\documentclass[twocolumn,prl,aps]{revtex4}
\usepackage{graphicx}
\draft

\newcommand{\be}{\begin{equation}}
\newcommand{\ee}{\end{equation}}
\newcommand{\ba}{\begin{eqnarray}}
\newcommand{\ea}{\end{eqnarray}}
\newcommand{\ban}{\begin{eqnarray*}}
\newcommand{\ean}{\end{eqnarray*}}

\begin{document}

\title{\bigskip Bell's theorem and Bohr's principle that the measurement must be classical}
\author{Antoine Suarez}
\address {Center for Quantum Philosophy, P.O. Box 304, CH-8044 Zurich, Switzerland.  E-mail: suarez@leman.ch}
\date{February 23, 2002}

\begin{abstract}
In a recent paper Karl Hess and Walter Philipp claim that hidden
local variables cannot be ruled out. We argue that their claim is
only valid if one gives up Bohr's principle that the measuring
instruments must be classical, and this principle belongs to the
foundations of scientific knowledge: Therefore, nonlocal
influences can be considered demonstrated.
\end{abstract}

 \maketitle

Bell's theorem \cite{jb64} states that certain mathematical
inequalities (``Bell's inequalities'') can be considered a
criterion to distinguish between Einstein's local realism
\cite{epr} and Quantum Mechanics: Local realistic theories satisfy
the inequalities, whereas Quantum Mechanics violates them.
Experiments conducted in the last two decades demonstrate such a
violation and the obtained results are in agreement with the
quantum mechanical predictions \cite{exp}. This fact has led to
the today widespread conviction that in nature there are nonlocal
influences acting faster than light \cite{jb64}, even though we
cannot use such ``Bell influences'' for faster than light
communication \cite{eb78}.

Nevertheless, Karl Hess and Walter Philipp have recently argued
that Bell theorem's proof is flawed. They propose a new local
model using so-called time-like correlated parameters, and claim
to prove that their extended space of local hidden variables does
permit derivation of the quantum predictions and is consistent
with all known experiments \cite{hp}. The argument has been
referred to as ``exorcising Einstein's spooks'' \cite{pb}.

In the following we first summarize the Hess-Philipp argument, and
then show that it should be considered an invitation to take
seriously Bohr's distinction between ``quantum object'' and
``classical measuring instrument'' \cite{nb}, rather than an
``exorcism'' of nonlocal influences.

The line of the Hess-Philipp argument is the following:

In his proof John Bell introduces an asymmetry in describing the
spin properties of the particles and the properties of the
measuring equipment. The particle's properties are described by
large sets $\Lambda$ of parameters. By contrast, the measurement
apparatus is described by a vector of the Euclidean space (the
settings), thus assuming in fact Bohr's postulate that the
measurement must be classical \cite{nb}. However the measurement
apparatus must itself in some form contain particles that, if one
wants to be self consistent, also need to be described by large
sets of parameters that are related to the settings.

Suppose the physicist at the analyzing station A chooses randomly
the setting {\bf a} for his measuring instrument, and the
physicist at the station B chooses randomly the setting {\bf b}
for his one. The settings {\bf a} and {\bf b} are obviously
uncorrelated. However, Hess and Philipp assume that the outcomes
at each station are not determined by the source $\Lambda$
(hidden) variables \emph{and} the settings {\bf a} and {\bf b},
but by the $\Lambda$ (hidden) variables \emph{and} certain
(hidden) parameters $a_1,....a_N$ (related to the setting {\bf a})
in station A, and $b_1,...b_N$ (related to the setting {\bf b})in
station B. Moreover, they assume that the parameters $a_1,....a_N$
are time-like correlated with the parameters $b_1,...b_N$, the
same way as the times indicated by a computer clock in New-York
are time-like correlated with those indicated by another clock in
Geneva. Under these assumptions the model clearly becomes a local
realistic description.

The last stone in the argument consists in showing that such a
local model permits (applying some involved mathematics) to derive
the correlations between the outcomes at A and B predicted by
Quantum Mechanics.

Actually the mathematical model is more general than described
above, and does not only encompass parameter sets that can be
labeled with the natural numbers. Nevertheless this is not
relevant for our discussion below. Note also that by setting in
motion the clock in Geneva one additionally introduces a Lorentz
transformation, but this does not break the times correlations
between the clocks. Therefore, experiments with moving apparatuses
\cite{asvs, as97, szgs} would not escape the Hess-Philipp argument
either, providing it holds.

In summary, the main assumption of Hess and Philipp is that
choosing at random the settings {\bf a} and {\bf b} at the arrival
of the particles into the stations does not ensure at all that the
space like separated measurements that determine the outcomes at A
and B are uncorrelated; they still may have time-related
correlations like two setting dependent clocks. But, as they
themselves acknowledge, this assumption is only valid if one gives
up Bohr's principle that the measurement is classical \cite{nb}.

So, strictly speaking, what Hess and Philipp really show is that
Bohr's distinction between "quantum object" and "classical
measuring instruments" \cite{nb} is a basic assumption in Bell's
proof, so that if one renounces to it, the proof fails, as it
fails if one questions that the physicist is capable of performing
free-willed choices. This clearly means also that after John
Bell's proof in 1964 a huge number of physicists have taken Bohr's
principle for the most obvious thing, from those who have
established versions of Bell's theorem and/or performed Bell
experiments, to those who have proposed local realistic models
(other than the Hess-Philipp's one) to explain the results of
Bell's experiments.

That so many distinguished physicists (as well localists as
nonlocalists) have so easily overlooked that Bell's proof bases on
Bohr's principle is a sign that this principle is somehow much
more natural than one uses to declare. Indeed that physics, also
quantum physics and even all scientific knowledge, begins with
classical observations and ends with classical observations can
hardly be denied. At the beginning we introduce classical
observable properties (time, length, mass, direction, charge,
etc.) we can freely vary acting upon. And at the end we register
classical observable events: the particle reaches one detector or
the other. Only thereafter, we conclude on quantum (classically
inexplicable) behavior: observing for instance interferences, we
deduce that a particle which can reach the detectors by two
alternative paths produces its outcome taking account at once of
information concerning both paths. Therefore, the hidden $\Lambda$
variables have to be defined with relation to the properties we
can freely control, i.e. the settings of the measuring
instruments.

This is especially clear in Bell experiments using Franson-type
interferometers exhibiting a long path and a short one
\cite{szgs}. Here the settings are the lengths $l$ and $s$ of the
long respectively short paths, and more precisely the phase-length
differences $\delta=l-s$ determining the phase parameters. Suppose
you would like to build a local hidden variables theory to explain
these experiments. The $\Lambda$ programs, presumed to be hidden
in the particles when they leave the source, cannot be
characterized otherwise that with relation to the path-length's
differences we set. These programs would consist in strings of the
form $\{\delta_1+, \delta_2-, \delta_3+, \delta_4-,...\}$
containing \emph{all} the possible path-length's differences
$\delta_i$ the physicists can choose, and the particles would meet
when they travel the interferometer, each $\delta_i$ having
assigned either value $+$ (the particle will undergo transmission
at the monitored beam-splitter) or $-$ (the particle will undergo
reflection at the monitored beam-splitter). If the programs are so
defined, then Bell's theorem holds.

This view of things leads to the conclusion that in the
Hess-Philipp model there is no measurement at all, for the
relevant variables $a_1....a_N$ and $b_1...b_N$ actually escape
the physicist's control.

To finish our discussion we would like to stress two points:

1. If one accepts Bohr's principle and defines the hidden
variables with relation to the possible setting choices the
physicists can make at the measuring stations, then any hidden
variable model rests on the assumption that the particle carries a
program containing all possible settings, all possible physicists
of all possible times may choose. This is for my taste a monstrous
idea far more difficult to swallow than the quantum mechanical
assumption, that is, each particle decides about the outcome in
arriving at the measuring apparatus though taking account of
nonlocal information. In this sense the Hess-Philipp paper helps
us to look at Nonlocality as something natural, even without
having to wait for violation of Bell's inequalities or other
locality criteria.

2. John Bell immensely contributed to increase our interest for
two quantum mechanical questions: Nonlocality and Measurement.
Regarding Nonlocality he was keen to know whether it is possible
to harmonize a time ordered causal description with Einstein's
relativity. In order to decide this question we have proposed
experiments with moving beam-splitters \cite{asvs, as00}. This
experiments have recently been done \cite{szgs} demonstrating a
new astonishing feature of quantum correlations: they escape
description in terms of ``before'' and ``after'' by means of any
set of real clocks, are brought about without relation to any real
timing; there is no real time ordering behind quantum causality
\cite{as01}. This means that at the fundamental level it is
impossible to unify Quantum Mechanics and Relativity, though this
has no observable consequence \cite{as01}. In this sense one of
the quantum conundrums Bell mainly bother about has been solved.
Regarding Measurement Bell was irritated by Bohr's division of the
world in classical and quantum, which he considered fuzzy and
unworthy of a precise theory \cite{jb90}. Ironically, it now
appears that if one gets rid of this division one gets rid of
Bell's theorem too, and also any physics, if we might say. So, the
sound and cheapest solution would rather consist in maintaining
both, Bohr's principle and Nonlocality. Admittedly, the question
of where we draw the line between quantum and classical or, in
Wheeler's wording, when a process of amplification becomes
irreversible and produces a registered phenomenon (i.e. when does
a detector actually click) \cite{jw}, remains a mystery still to
elucidate.

In conclusion, the Hess-Philipp's paper doesn't invalidate the
proofs of nonlocal influences but invite us to reflect more in
depth about Bohr's principle. Experiments with space-like
apparatuses in motion seem to have completed the characterization
of quantum Nonlocality. Apparently, the interesting fundamental
question to investigate now is that of when the things we see and
control emerge from the invisible quantum world.

I would like to thank Karl Hess for clarifying remarks, and the
Odier Foundation of Psycho-physics and the L\'eman Foundation for
financial support.

{}

\end{document}